# Billion-Fold Enhancement of Room-Temperature Ionic Conductivity in h-$R$MnO$_3$/YSZ Heterostructures via Electric-Field-Assisted Oxygen Deficiency Engineering

Detian Yang,[1] Yaohua Liu,[2] Liang Dai,[1] Zhihang Xu,[1] Xiaoshan Xu[3,4]*

[1] *Shanghai Key Laboratory of High Temperature Superconductors, Department of Physics, Shanghai University, Shanghai 200444, China*
[2] *Second Target Station, Oak Ridge National Laboratory, Oak Ridge, Tennessee 37830, USA*
[3] *Department of Physics and Astronomy, University of Nebraska, Lincoln, Nebraska 68588, USA44*
[4] *Nebraska Center for Materials and Nanoscience, University of Nebraska, Lincoln, Nebraska 68588, USA*

Oxide heterostructures provide versatile platforms for manipulating electronic and ionic conductive states. In this study, we demonstrate a remarkable billion-fold enhancement in room-temperature ionic conductivity within h-$R$MnO$_3$/YSZ heterostructures, achieved through electric-field-assisted oxygen deficiency engineering. This enhancement is closely linked to substantial oxygen depletion in YSZ and is tunable by varying the thickness of the h-$R$MnO$_3$ film layer and the applied voltage bias. Our findings underscore the critical importance of interfacial design and vacancy control in enhancing ionic transport capabilities, paving the way for advanced applications in low-temperature energy harvesting, storage, and conversion technologies.

## I. INTRODUCTION

Oxide interfaces and heterostructures have garnered significant interest due to their potential to host unique physical properties that are absent in the bulk phases of the constituent materials [1–5]. These engineered interfaces allow for boosted electronic and ionic conductivity, opening pathways to novel functionalities in complex oxide systems. Prominent examples include interfacial superconductivity between insulating La$_2$CuO$_4$ and metallic (La,Sr)$_2$CuO$_4$ [6], and the formation of a two-dimensional electron gas at the interface between insulating LaAlO$_3$ and SrTiO$_3$ [7]. Similarly, ionic conductivity can be enhanced by interfacial and heterostructural engineering [1,5,8], which holds promising applications in energy-related technologies, such as solid oxide fuel cells (SOFCs), batteries, and chemical filters and sensors [9]. By intentionally designing heterostructures with increased interface density, symmetric boundary conditions, and space-charge regions, significant gains in ionic conductivity have been achieved. Examples include heterostructures like YSZ/MgO [10,11], CSZ/Al$_2$O$_3$ [12], YSZ/Y$_2$O$_3$ [13] and SDC/YSZ [14], where ionic conductivity is substantially improved in comparison with single-phase ionic conductors.

A primary application of ionic conductors, such as YSZ [15] and YSC [16], is their role as electrolytes in SOFCs. However, a major limitation is the high operational temperature, typically between 800°C and 1000°C, needed to achieve sufficient ionic conductivity. Efforts over the past three decades have focused on reducing these operating temperatures to 500–700°C through the development of new materials and heterostructures with improved ionic conductivity at lower temperatures [17–19].

Recent studies recognize high oxygen vacancy concentrations near surfaces and interfaces as essential for these enhanced effects. Increased oxygen vacancies in thin film heterostructures could not only reshape electronic properties—such as enabling metal-insulator transitions [20] and resistive switching [21] —but also effectively improve catalytic activity [22] and ionic conductivity [23]. For instance, trilayer SrTiO$_3$/YSZ/SrTiO$_3$ [24–26] structures exhibit an ionic conductivity enhancement of up to eight orders of magnitude near room temperature, attributed primarily to interfacial strain and space-charge effects that modify defect chemistry and promote oxygen vacancy formation and mobility [27]. Also, in In$_2$O$_3$/YSZ heterostructures, which combine an electronic semiconductor film with an ionic conductor, it was demonstrated that small in-plane fields can induce interfacial oxygen transfer and enhance the film's electrical conduction by two orders of magnitude [28,29].

Hole-doped perovskite manganites exhibit properties such as colossal/giant magnetoresistance [30], charge/orbital orderings [31] and metal/superconductor-insulator transition [32] due to double-exchange interactions among manganese ions with mixed valences. With their high catalytic activity for oxygen reduction reactions and high mixed ionic and electronic conductivity, perovskite manganites like $La_{1-x}Sr_xMnO_3$ are promising candidates as cathode materials for SOFCs [33–35]. However, their hexagonal counterparts—doped h-$R$MnO$_3$($R$=Ho-Lu) [36], such as h-$Yb_{1-x}Ca_xMnO_3$ (x = 0, 0.05, 0.10) [37], h-$Yb_{1-x}Mg_xMnO_3$ (x=0.00, 0.05) [38] and h-$Lu_{1-x}Ca_xMnO_3$ (x = 0.1, 0.3, 0.5) [39] —show suppressed double exchange interactions and electronic conductivity. This is likely due to the symmetry-adapted spin-frustrated structures in hexagonal manganites, which favour superexchange interactions and prefer to achieve valence balance partly through increased oxygen vacancy formation, rather than just altering the valence of manganese ions [40,41].

In this work, by studying doped and undoped h-$R$MnO$_3$ ($R$= Y, $Lu_{1-x}Ca_x$, x=0, 0.1, 0.2, 0.3, 0.4, 0.5) films on YSZ substrates, we demonstrate a billion-fold enhancement in room-temperature ionic conductivity achieved through electric-field-assisted oxygen deficiency engineering. This conductivity boost is directly tied to the presence of an h-$R$MnO$_3$ film layer, with a critical thickness of approximately 3-4 nm below which the enhancement become negligible. The improvement is also sensitive to voltage bias yet remains largely independent of the specific rare-earth element within the h-$R$MnO$_3$ layer as well as doping effect. Our findings highlight an effective strategy for enhancing oxygen vacancy mobility in YSZ at temperatures below 500 K, underscoring the critical role of oxygen vacancy engineering at oxide interfaces. This work offers valuable insights into the design of high-performance ionic conductors for energy conversion and storage applications.

## II. EXPERIMENTS

**Sample preparation**. h-$Lu_{1-x}Ca_xMnO_3$(001)/h-$YMnO_3$(001)/YSZ(111)(x=0, 0.1, 0.2, 0.3, 0.4, 0.5), h-$YMnO_3$(001)/YSZ(111) and h-$LMnO_3$(001)/YSZ(111) epitaxial thin films were grown using pulsed laser deposition (PLD) in an oxygen background pressure of 80 mTorr. All films were deposited at 1020 K and subsequently annealed at 1100 K for 10–30 minutes in an oxygen atmosphere. The growth temperature was monitored with two pyrometers: one installed within the PLD chamber, and the other positioned externally for cross-verification. A KrF excimer laser of the wavelength 248 nm was employed to ablate the targets with a pulse energy of 110 mJ and a repetition rate of 2 Hz. The growth processes were in-situ monitored by a reflection high energy electron diffraction (RHEED) system.

**Structure characterization**. Structural characterization of the films was conducted using the out-of-plane $\theta$-$2\theta$ x-ray diffraction (XRD) and x-ray reflectivity (XRR), performed on a Rigaku SmartLab x-ray diffractometer (copper K-α source, X-ray wavelength 1.5406 Å); the film thicknesses were derived from the XRR data.

**Transport Measurements**. For electrical transport measurements, low-temperature (20 K < T < 300 K) characterization was conducted using a custom-built cryostat system with a helium cooler. High-temperature (300K<T<500K) measurements were performed using a Lakeshore TTP4 Probe Station, which included a Wat-231S2 camera and a Preiffer HiPace 80 pump. Current-voltage (I-V) and temperature-dependent resistance (R-T) measurements were conducted using a Keithley 236 SourceMeter with the Van der Pauw method. Unless otherwise specified, the heating and cooling rates were maintained at less than 2 K/min, and the voltage bias in the R-T measurements was set to 100 Vhigh performance.

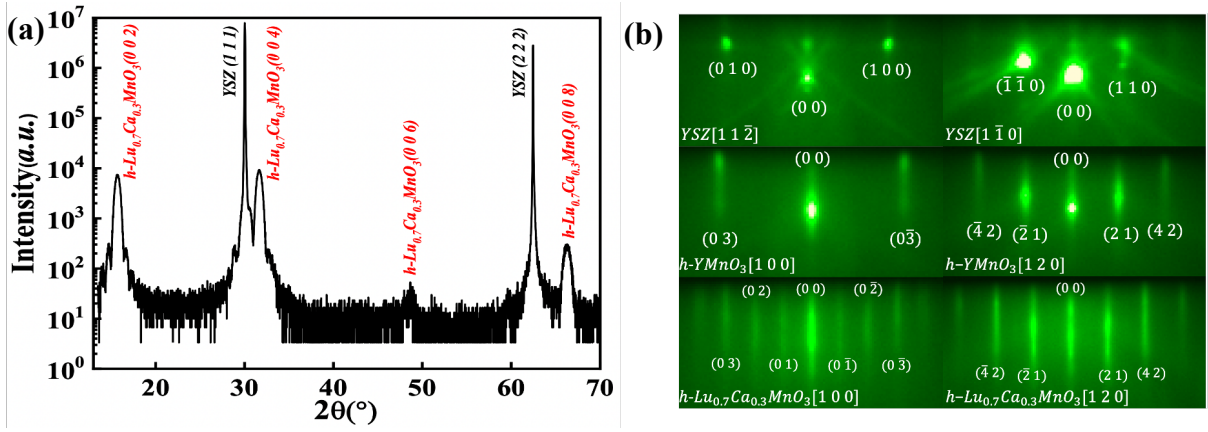

**Fig.1.** (a) $\theta - 2\theta$ scan of XRD for a 13 nm h-Lu$_{0.7}$Ca$_{0.3}$MnO$_3$ (001)/h-YMnO$_3$(001)/YSZ (111) sample. (b) Typical RHEED patterns of the 12 nm Lu$_{0.7}$Ca$_{0.3}$MnO$_3$(bottom panel), 1 nm buffer layer YMnO$_3$(middle panel) and the YSZ substrate (top panel) along two perpendicular in-plane directions: YSZ [11-2] and YSZ [1-10].

## III. RESULTS AND DISCUSSION

### A. Growth and structural characterization

The (001)-oriented hexagonal manganite thin films could be epitaxially deposited on YSZ (111) substrates despite their incoherent interfaces formed due to different symmetries and incommensurable lattice parameters [42,43]. Previous studies have shown that using a high-quality hexagonal manganite buffer layer of higher crystal quality and better flatness can improve the crystallinity and flatness of subsequent thick h-YMnO$_3$ thin films grown on YSZ substrates [44]. To stabilize Lu$_{1-x}$Ca$_x$MnO$_3$ in its hexagonal structure on cubic YSZ substrates, we first deposit a 1 nm h-YMnO$_3$ buffer layer. This buffer layer provides a crystal surface with the desired symmetry and structure, compatible with the hexagonal phase of Lu$_{1-x}$Ca$_x$MnO$_3$ and induces tensile strain in the film (see Fig. S1(e)).

The hexagonal structure of h-$R$MnO$_3$ thin films is illustrated by the out-of-plane $\theta - 2\theta$ XRD scan of Lu$_{0.7}$Ca$_{0.3}$MnO$_3$ in Fig. 1(a), where no impurity phases are observed (see Fig. S1 for additional XRD data of RMnO$_3$ films). Clear Laue oscillations further indicate excellent crystallinity, as confirmed by the rocking curve in Fig. S1(d) with a full width at half maximum (FWHM) of only 0.13°, highlighting the high structural quality of the films.

The epitaxial relationship among the Lu$_{1-x}$Ca$_x$MnO$_3$ films, h-YMnO$_3$ buffer layers, and YSZ substrates is summarized in Fig.S1(g), derived from the RHEED patterns of h-Lu$_{0.7}$Ca$_{0.3}$MnO$_3$ shown in Fig. 1(b) and the φ-scan in Fig. S1(h). Both the RHEED patterns for the h-YMnO$_3$ buffer layer and h-Lu$_{0.7}$Ca$_{0.3}$MnO$_3$ films are indexed based on the reciprocal primitive lattice of h-Lu$_{0.7}$Ca$_{0.3}$MnO$_3$, with a *P6$_3$cm* symmetry group. In the typical φ-scans, six-fold and three-fold symmetry in the film and substrate plane are demonstrated by the six and three equivalent peaks corresponding to the (0 3 8) reflection of Lu$_{0.7}$Ca$_{0.3}$MnO$_3$ and the (0 4 4) reflection of the YSZ substrate, respectively, which confirms the epitaxial alignment.

Thermal stability of $Lu_{0.7}Ca_{0.3}MnO_3$ was examined by sequential thermal annealing and XRD characteriaztion, as shown in Fig. S2. The results indicate that $Lu_{0.7}Ca_{0.3}MnO_3$ retains stability up to 950 K, above which decomposition occurs. This stability under high temperature is crucial for practical applications where thermal endurance is required.

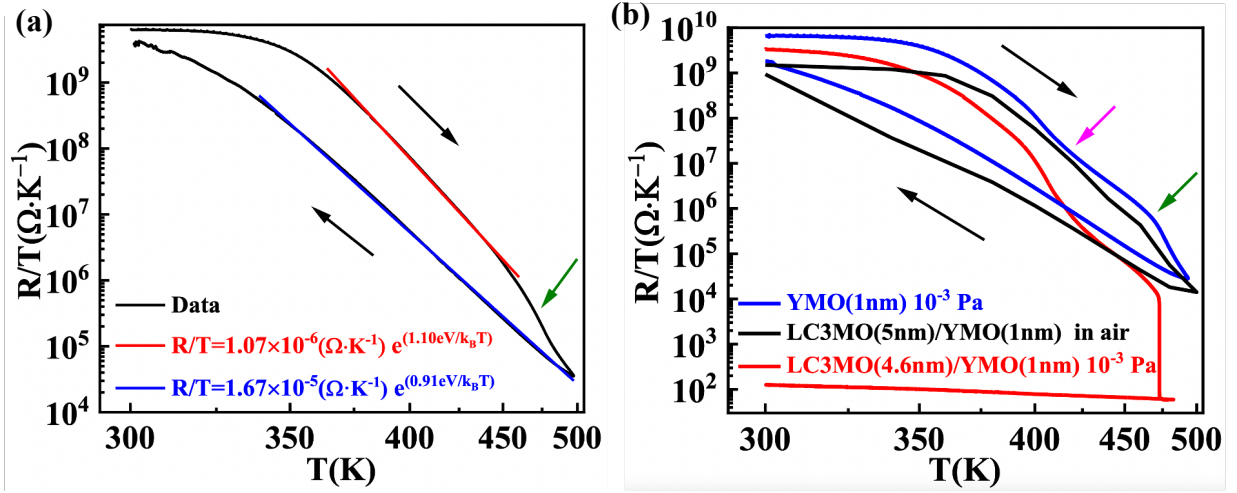

**Fig.2. (a)** R-T curve (black line) of an annealed YSZ substrate in the vacuum of $10^{-3}$ Pa, with red and blue lines representing linear fits for the heating and cooling branches, respectively. **(b)** R-T curves of a 1nm h-YMnO$_3$ (001)/YSZ(111) sample (blue line), a 6 nm h-Lu$_{0.7}$Ca$_{0.3}$MnO$_3$(001)/h-YMnO$_3$(001)/YSZ(111) sample (black line, measured in air) and a 5.6 nm h-Lu$_{0.7}$Ca$_{0.3}$MnO$_3$(001)/ YMnO$_3$(001)/YSZ(111) sample (red line, measured in vacuum at $10^{-3}$ Pa). LC3MO and YMO denote h-Lu$_{0.7}$Ca$_{0.3}$MnO$_3$ and h-YMnO$_3$, respectively. Measurements proceeded first with heating, then cooling, as indicated by black arrows. The R/T and T axes are in logarithmic and reciprocal scales, respectively. See main text for additional details.

## B. R-T characterization and sharp resistance drop

To evaluate the conductivity contribution of YSZ substrates in h-$R$MnO$_3$/YSZ heterostructures, we measured the R–T curve of a YSZ substrate subjected to similar annealing procedures as the growth process of the h-$R$MnO$_3$ films. Figure 2(a) displays its R–T curve, showing two distinct, approximately linear regions that fit well with the Arrhenius relationship [45,46]: one in the heating range (370–465 K) and the other in the cooling range (500–350 K). The linear fits yield activation energies of $E_a$ = 1.10 eV for heating and $E_a$ = 0.91 eV for cooling, matching reported activation energy values of YSZ [47–49]. A faster resistance drop (olive arrow) in the 465–500 K range suggests a density increase in free oxygen vacancies due to combined effects of voltage bias and thermal energy, which decreases the activation energy from 1.10 to 0.91eV as observed. At lower temperatures, the R–T curve deviates from the Arrhenius relation, likely due to a decreased density of free oxygen vacancies [45].

In the $R$MnO$_3$/YSZ heterostructure, the overall resistance reflects a parallel combination of the $R$MnO$_3$ film resistance and the ionic resistance of the YSZ substrate. Figure 2(b) presents the R–T curves for three thin-film samples. The R–T behaviors of both the 6 nm Lu$_{0.7}$Ca$_{0.3}$MnO$_3$/YMnO$_3$/YSZ thin film sample measured in air and the 1 nm YMnO$_3$/YSZ film sample measured in vacuum resemble the trend and magnitude observed in the YSZ substrate, as shown in Fig. 2(a), highlighting the dominant role of YSZ ionic conductivity in these samples. Notably, both samples exhibit a clear dip around 420 K (indicated by the magenta arrow) in the heating curves under vacuum conditions. Additionally, in the 5.6 nm Lu$_{0.7}$Ca$_{0.3}$MnO$_3$/YMnO$_3$ film sample measured in vacuum, a sharp resistance drop occurs around 475 K, reducing the resistance by nearly two orders of magnitude. This resistance jump is consistent with downturns observed in the R–T curves of both the YMnO$_3$/YSZ and YSZ samples (see the olive arrow), suggesting a common underlying mechanism amplified within the Lu$_{0.7}$Ca$_{0.3}$MnO$_3$/YMnO$_3$/YSZ heterostructure.

The intermediate region between the dip and downturn/jump points (420–475 K) in these two *R–T* curves, both measured in vacuum, aligns well with an Arrhenius-type behavior (see Figures S4(a.1) and S4(d.1)) with fitted activation energies of 1.11 eV and 1.12 eV, respectively, closely matching the YSZ substrate's activation energy 1.10 eV extraced in the same range, further underscoring the dominant influence of YSZ's ionic conductivity. Interestingly, in the cooling process, the 5.6 nm Lu$_{0.7}$Ca$_{0.3}$MnO$_3$/YMnO$_3$/YSZ sample maintains a low resistance up to room temperature, exhibiting only a minor resistance increase. XRD data indicates that the heterostructure's crystal structure remains stable, with only a slight broadening in the rocking curves of the Lu$_{0.7}$Ca$_{0.3}$MnO$_3$ film (see Figure S5). Similar to the YSZ substrate, cooling R–T curves of both the 1 nm YMnO$_3$/YSZ and 5.6 nm Lu$_{0.7}$Ca$_{0.3}$MnO$_3$/YMnO$_3$/YSZ samples can be fit to two Arrhenius-type relations in the temperature ranges 350 K < T < 500 K and 300 K < T < 350 K (see Figures S4(a.2) and S4(d.2)). In the range of 350 K < T < 500 K, activation energies of 0.86 eV and 57.8 meV are obtained for these films, significantly reduced from the 9.1 eV of the YSZ substrate by factors of 1.06 and 28.9, respectively, highlighting the impact of heterostructuring on activation energy.

Upon cooling to 300 K, the conductivity enhancement for the 5.6 nm Lu$_{0.7}$Ca$_{0.3}$MnO$_3$/YMnO$_3$/YSZ sample reaches approximately $10^7$ times its initial value prior to R–T measurement (and more than $10^9$ times in a 10 nm Lu$_{0.8}$Ca$_{0.2}$MnO$_3$/YMnO$_3$/YSZ sample shown later in Fig.5). Comparison between R–T curves for the 6 nm Lu$_{0.7}$Ca$_{0.3}$MnO$_3$/YMnO$_3$/YSZ sample measured in air and the 5.6 nm Lu$_{0.7}$Ca$_{0.3}$MnO$_3$/YMnO$_3$/YSZ sample measured in vacuum underscores the essential role of the vacuum environment in enabling the resistance jump. Furthermore, during the vacuum heating process for the 5.6 nm Lu$_{0.7}$Ca$_{0.3}$MnO$_3$/YMnO$_3$/YSZ sample, the YSZ substrate exhibits a pronounced darkening to jet black (Figure S6) after the dip at 420 K, indicative of significant oxygen loss [50–53]. This oxygen depletion is not observed in other samples, establishing a strong correlation between oxygen loss and the resistance jump. It also underscores the crucial influence of both the manganite film layer and the vacuum environment in achieving these effects. Notably, this correlation between oxygen loss and resistance jump consistently appears in all samples displaying a giant conductivity enhancement (a factor of larger than $10^5$) in this study.

### C. A qualitative model for giant conductivity enhancement

Based on the observations discussed above, we propose that the substantial conductivity enhancement arises from an increase in oxygen vacancy channels caused by electric-field-induced oxygen depletion. Previous studies have shown that voltage differences can drive oxygen ion transfer across interfaces in ln$_2$O$_3$/YSZ heterostructures [28,29]. The faster resistance drops observed between 465 and 500 K(indicated by the olive arrow in Fig. 2) suggest that a high voltage bias of 100 V may enable direct oxygen extraction from the YSZ substrate, assisted by thermal energy and the Pt electrode interface. Presumably, the removed oxygens are then released into the vacuum chamber as O$_2$.

As in our work, h-*R*MnO$_3$ films grown on YSZ substrates generally experience tensile strain [54,55]. The strong coupling between interfacial strain and oxygen vacancies can efficiently stabilize oxygen vacancies and multiple experimental and computational reports have proposed that tensile strain decreases the formation energy of oxygen vacancy and increases its concentration near the interfaces [56–60]. This increase in interfacial oxygen vacancies can be further amplified by the space-charge region, particularly within YSZ [61–65]. Obviously, as indicated in Fig.3, the resultant interfacial region full of oxygen vacancies can more effectively transfer $O^{2-}$ when a high voltage bias is applied. Note that the hexagonal close-packed (hcp) oxygen sublattice in the hexagonal manganite film layer could provide a dense array of oxygen sites and efficient pathways for ion migration, potentially lowering activation energy and stabilizing oxygen vacancies. This feature, known to enhance ionic conductivity in other fast ionic conductors like Beta-Alumina and $La_2NiO_{4+\delta}$, [66,67] could similarly contribute to both ionic and electronic conductivity here under the applied voltage. Consequently, significant oxygen removal creates additional free oxygen vacancies in both the YSZ substrate and the manganite film layer, collectively leading to the observed resistance reduction.

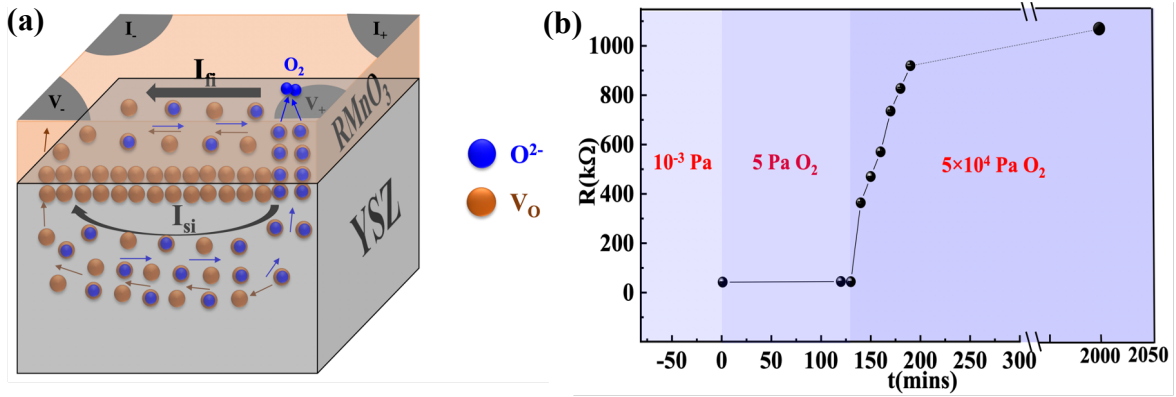

**Fig.3.** (a) Schematic model illustrating voltage-assisted oxygen extraction and enhanced ionic conductivity via oxygen vacancy channels formed within the film and YSZ substrate. Brown and blue spheres represent oxygen vacancies ($V_o$) and $O^{2-}$ ions, with corresponding arrows showing their diffusion directions. $I_{fi}$ and $I_{si}$ denote ionic currents in the film and substrate, respectively. Pt electrodes are shown in grey at the corners of the film, and $V_+$($V_-$) and $I_+$($I_-$) indicate positive (negative) voltage bias and current terminals. (b) Conductivity decay over time at 300 K in an oxygen atmosphere for the 5.6 nm $Lu_{0.7}Ca_{0.3}MnO_3$/$YMnO_3$/YSZ sample after the R-T measurement in Fig. 2(b). Time zero marks the introduction of 5 Pa $O_2$ into the measurement chamber.

To further validate the role of oxygen vacancies and confirm the ionic nature of the low-resistance state, we exposed the 5.6 nm $Lu_{0.7}Ca_{0.3}MnO_3$/$YMnO_3$/YSZ sample to oxygen at 300 K after the *R-T* measurement (Fig. 3(b)). Resistance measurements over time (extracted from *I-U* curves in Fig. S7) reveal that, in 5 Pa $O_2$, the resistance remained low even after 130 minutes. However, under a higher pressure of $5\times10^4$ Pa $O_2$, resistance gradually increased, resulting in a 30-fold reduction in conductivity over 30 hours. The resistance eventually returned to the $10^9$–$10^{11}$ Ω range after 3-hour annealing at 500 K in air or exposure to air for two weeks, indicating a decrease in charge carrier density as oxygen vacancies recombined with oxygen. This supports the preeminent role of oxygen vacancies in the conductivity enhancement.

### D. Tunability by film thickness and voltage bias

In Fig. 2(b), we observe that a 1 nm $YMnO_3$ layer does not induce a resistance jump in the $YMnO_3$/YSZ heterostructure, indicating that the thickness of the $R$MnO$_3$ layer may be influential. To further examine this, we recorded R-T curves for $Lu_{0.7}Ca_{0.3}MnO_3$/$YMnO_3$/YSZ heterostructures at varying film thicknesses, as shown in Fig. 4(a). The resistance jump appears only when the film thickness reaches 3.8 nm, suggesting a threshold below which the manganite layer does not adequately facilitate oxygen vacancy formation for conductivity enhancement. This threshold effect likely reflects the need for a minimum number of oxygen sites to provide sufficient interfacial oxygen vacancies to extract oxygen from YSZ.

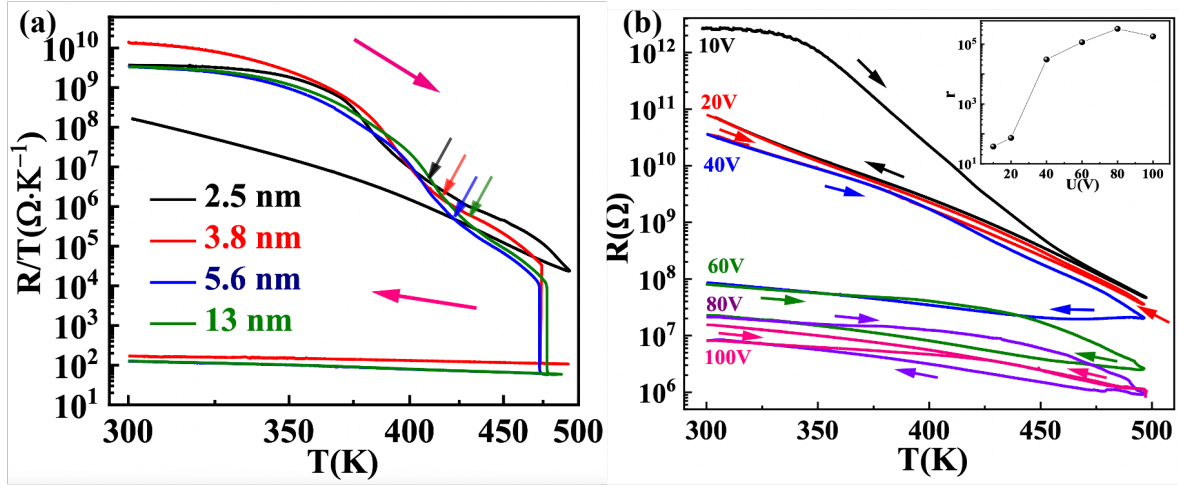

**Fig.4.** (a) R-T curves of h-Lu$_{0.7}$Ca$_{0.3}$MnO$_3$(001)/h-YMnO$_3$(001)/YSZ(111) heterostructures with varying film thicknesses. Pink arrows indicate measurement directions, and "dip" points are marked by arrows matching the colour of each curve. The R/T and T axes are presented in logarithmic and reciprocal scales, respectively. (b) R-T curves measured under sequential voltage biases: 10V (black), 20V (red), 40V (blue), 60V (olive), 80V (purple), and 100V (magenta). Measurement directions are indicated by arrows in corresponding colours. The inset shows the conductivity enhancement ratio $r$ at 300 K versus voltage bias $U$, derived from these R-T curves. r (U) is defined the conductivity ratio at the end of each run compared to the initial conductivity at 300K.

Notably, the temperature at which the resistance jump occurs remains consistent, around 475 ± 3K, regardless of thickness in the range of 3.8 to 13 nm. However, the "dip" temperatures, indicated by arrows matching the R-T curve colours, increase with the thicknesses, and no dips are observed in thicker samples (see Fig. S8), pointing to an interfacial origin for the dips. As with previous observations, the region between the dip and downturn or jump points reflects the ionic conductivity of YSZ, with Arrhenius fittings (Fig.S4(b.1)-(e.1)) yielding activation energies from 0.94 eV to 1.14 eV as the thickness increases.

As noted in section 3.2.2, each cooling R-T curve divides into two segments that fit well to the Arrhenius-type relationship. With the resistance jump, both segments show a significant drop in activation energy, by a factor of about 10-15, once the film thickness exceeds the critical thickness, as illustrated in Fig. S4(f.1) and (f.2). Intriguingly, the low activation energies (16-60meV) are close to those observed in La$_{1-x}$Ca$_x$MnO$_3$ (0≤x≤0.8) with mixed ionic and electronic conductivity [68].

As detailed in section 3.2.2, a voltage bias plays a crucial role in extracting oxygen from YSZ substrates, thereby significantly enhancing ionic conductivity. Fig.4(b) presents R-T measurements on a 5.6 nm Lu$_{0.7}$Ca$_{0.3}$MnO$_3$/YMnO$_3$/YSZ heterostructure across six voltage increments from 10 to 100 V, each run comprising both heating and cooling. Conductivity enhancement after each run is quantified by the ratio $r$, which compares the initial resistance at 300 K (before R-T measurements) to the resistance after each run at 300 K. As shown in the inset of Fig.4(b), $r$ increases with voltage bias $U$, saturating around $10^5$ at 80-100V, with a significant rise between 30-40 V. Assumingly, this threshold corresponds to the voltage required for efficient oxygen vacancy formation and provides sufficient kinetic energy for O$^{2-}$ ions to transfer across the interface.

Note that, unlike the single-run 100 V R-T curve in Fig.2(b), which resulted in a $10^7$-fold conductivity enhancement, the sequential lower-voltage runs achieved only a $10^5$-fold increase. This discrepancy suggests that oxygen vacancy channel formation is non-equilibrium and sensitive to the thermodynamic history, with an initial high voltage critical for achieving maximum conductivity enhancement.

### E. Universality of conductivity enhancement in h-$R$MnO$_3$/YSZ heterostructures

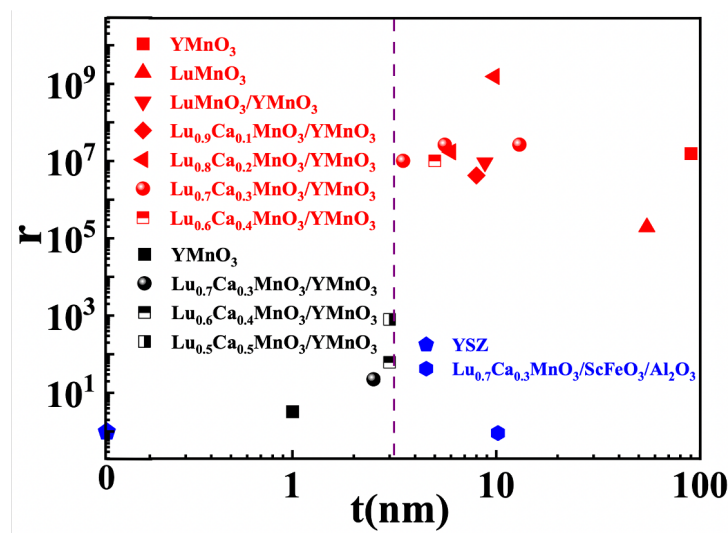

**Fig.5.** Thickness-dependent conductivity enhancement ratio ($r$) of various thin film heterostructures. Samples represented by red symbols on YSZ substrates show significant conductivity enhancement (>$10^5$), while those in black exhibit only minor improvement. For h-Lu$_{1-x}$Ca$_x$MnO$_3$/h-YMnO$_3$/YSZ (x=0, 0.1, 0.2, 0,3, 0.4, 0.5) samples, the YMnO$_3$ buffer layer thickness was fixed at 1nm. Reference data from a YSZ substrate (blue pentagon) and a h-Lu$_{0.7}$Ca$_{0.3}$MnO$_3$(001)/h-ScFeO$_3$(001)/Al$_2$O$_3$(001) heterostructure on $\alpha$-Al$_2$O$_3$ (blue hexagon) are also shown. The purple line indicates the critical thickness.

We have systematically explored the substantial ionic conductivity enhancement in h-Lu$_{0.7}$Ca$_{0.3}$MnO$_3$(001)/h-YMnO$_3$(001)/YSZ(111) thin film heterostructures. Here, we extend these findings across a range of h-$R$MnO$_3$(001)/YSZ(111) heterostructures, demonstrating that similar conductivity enhancement consistently arises once the h-$R$MnO$_3$ layer thickness surpasses a critical threshold of approximately 3-4 nm.

To verify this, we performed R-T measurements on various h-$R$MnO$_3$ thin films(see Fig.S8) and calculated the conductivity enhancement ratio ($r$) for each sample, plotting it against film thickness ($t$) in Fig. 5. Notably, a 10 nm h-Lu$_{0.8}$Ca$_{0.2}$MnO$_3$(001)/h-YMnO$_3$(001) film on YSZ (111) exhibited a billion-fold conductivity increase. Across all h-$R$MnO$_3$ thin films on YSZ substrates, resistance drops significantly when the thickness exceeds 3–4 nm, as marked by the purple line in Fig. 5. The consistency across various rare-earth ions supports the importance of the hexagonal-close-packed oxygen sublattice, rather than rare-earth ions, in forming oxygen vacancy channels that boost ionic conductivity in h-$R$MnO$_3$(001)/YSZ(111) structures. Clearly, such a critical thickness effect cannot be solely attributed to interfacial phenomena.

Furthermore, the similar $10^7$-fold conductivity enhancement found in 5–6 nm films with Ca doping (x = 0.2–0.4) in h-Lu$_{1-x}$Ca$_x$MnO$_3$/h-YMnO$_3$/YSZ indicates that hole doping levels minimally impact the enhancement, against a doping-induced electronic or mixed conductivity mechanism seen in perovskite La$_{1-x}$Ca$_x$MnO$_3$ [68]. In contrary, our model indicates that the considerable formation of oxygen vacancy channels in YSZ is essential to the conductivity enhancement, confirmed by the lack of conductivity enhancement in heterostructures lacking YSZ (e.g.h-Lu$_{0.7}$Ca$_{0.3}$MnO$_3$(001)/h-

ScFeO$_3$(001)/Al$_2$O$_3$(001)), reinforcing the essential role of YSZ in facilitating the conductivity improvements.

## IV. CONCLUSIONS

In summary, our work demonstrates a significant enhancement in room-temperature ionic conductivity within h-$R$MnO$_3$/YSZ heterostructures, tunable by both the thickness of the h-$R$MnO$_3$ film and applied voltage bias. Remarkably, a billion-fold increase in conductivity was observed in a 10 nm h-Lu$_{0.8}$Ca$_{0.2}$MnO$_3$(001)/h-YMnO$_3$(001) film on YSZ. This enhancement, closely associated with substantial oxygen depletion in YSZ, and its sensitivity to oxygen atmosphere, suggests that electric-field-assisted oxygen vacancy engineering has induced a dramatic increase in oxygen vacancy channels within the h-$R$MnO$_3$ /YSZ heterostructures.

These results highlight the importance of interfacial design and vacancy control in advancing ionic transport capabilities. The insights gained hold promise for optimizing ionic conductors, with wide-ranging applications in low-temperature solid oxide fuel cells, advanced batteries, and energy conversion technologies.

## ACKNOWLEDGEMENTS

The authors acknowledge the primary support from the National Science Foundation (NSF) through EPSCoR RII Track-1: Emergent Quantum Materials and Technologies (EQUATE), Award No.OIA-2044049. The research was performed in part in the Nebraska Nanoscale Facility: National Nanotechnology Coordinated Infrastructure and the Nebraska Center for Materials (NCMN) and a d Nanoscience and Nano-Engineering Research Core Facility (NERCF), which are supported by the NSF under Grant No. ECCS-2025298, and the Nebraska Research Initiative.

[1] E. Fabbri, D. Pergolesi, and E. Traversa, Sci Technol Adv Mater **11**, 054503 (2010).
[2] J. A. Sulpizio, S. Ilani, P. Irvin, and J. Levy, Annu Rev Mater Res **44**, 117 (2014).
[3] P. Zubko, S. Gariglio, M. Gabay, P. Ghosez, and J.-M. Triscone, Annu Rev Condens Matter Phys **2**, 141 (2011).
[4] Z. Huang, Ariando, X. Renshaw Wang, A. Rusydi, J. Chen, H. Yang, and T. Venkatesan, Advanced Materials **30**, (2018).
[5] X. Guo and J. Maier, Advanced Materials **21**, 2619 (2009).
[6] A. Gozar, G. Logvenov, L. F. Kourkoutis, A. T. Bollinger, L. A. Giannuzzi, D. A. Muller, and I. Bozovic, Nature **455**, 782 (2008).
[7] S. Thiel, G. Hammerl, A. Schmehl, C. W. Schneider, and J. Mannhart, Science (1979) **313**, 1942 (2006).
[8] G. Knöner, K. Reimann, R. Röwer, U. Södervall, and H.-E. Schaefer, Proceedings of the National Academy of Sciences **100**, 3870 (2003).
[9] N. Sata, K. Eberman, K. Eberl, and J. Maier, Nature **408**, 946 (2000).
[10] I. Kosacki, C. M. Rouleau, P. F. Becher, J. Bentley, and D. H. Lowndes, Electrochemical and Solid-State Letters **7**, A459 (2004).

[11] M. Sillassen, P. Eklund, N. Pryds, E. Johnson, U. Helmersson, and J. Bøttiger, Adv Funct Mater **20**, 2071 (2010).
[12] A. PETERS, C. KORTE, D. HESSE, N. ZAKHAROV, and J. JANEK, Solid State Ion **178**, 67 (2007).
[13] C. Korte, A. Peters, J. Janek, D. Hesse, and N. Zakharov, Physical Chemistry Chemical Physics **10**, 4623 (2008).
[14] S. Sanna, V. Esposito, D. Pergolesi, A. Orsini, A. Tebano, S. Licoccia, G. Balestrino, and E. Traversa, Adv Funct Mater **19**, 1713 (2009).
[15] P. Vinchhi, M. Khandla, K. Chaudhary, and R. Pati, Inorg Chem Commun **152**, 110724 (2023).
[16] N. Mahato, A. Gupta, and K. Balani, Nanomaterials and Energy **1**, 27 (2012).
[17] J. A. Kilner and M. Burriel, Annu Rev Mater Res **44**, 365 (2014).
[18] Z. Gao, L. V. Mogni, E. C. Miller, J. G. Railsback, and S. A. Barnett, Energy Environ Sci **9**, 1602 (2016).
[19] Z. Zakaria, S. H. Abu Hassan, N. Shaari, A. Z. Yahaya, and Y. Boon Kar, Int J Energy Res **44**, 631 (2020).
[20] J. Jeong, N. Aetukuri, T. Graf, T. D. Schladt, M. G. Samant, and S. S. P. Parkin, Science (1979) **339**, 1402 (2013).
[21] Y. Yang, S. Choi, and W. Lu, Nano Lett **13**, 2908 (2013).
[22] H. Jeen, Z. Bi, W. S. Choi, M. F. Chisholm, C. A. Bridges, M. P. Paranthaman, and H. N. Lee, Advanced Materials **25**, 6459 (2013).
[23] H. Cai, C. Xia, X. Wang, W. Dong, H. Xiao, D. Zheng, H. Wang, and B. Wang, ACS Appl Energy Mater **5**, 11122 (2022).
[24] J. Garcia-Barriocanal, A. Rivera-Calzada, M. Varela, Z. Sefrioui, E. Iborra, C. Leon, S. J. Pennycook, and J. Santamaria, Science (1979) **321**, 676 (2008).
[25] J. García-Barriocanal, A. Rivera-Calzada, M. Varela, Z. Sefrioui, E. Iborra, C. Leon, S. J. Pennycook, and J. Santamaría, Science (1979) **324**, 465 (2009).
[26] X. Guo, Science (1979) **324**, 465 (2009).
[27] T. J. Pennycook, M. J. Beck, K. Varga, M. Varela, S. J. Pennycook, and S. T. Pantelides, Phys Rev Lett **104**, 115901 (2010).
[28] B. W. Veal, S. K. Kim, P. Zapol, H. Iddir, P. M. Baldo, and J. A. Eastman, Nat Commun **7**, 11892 (2016).
[29] B. W. Veal and J. A. Eastman, APL Mater **5**, (2017).
[30] S. Jin, T. H. Tiefel, M. McCormack, R. A. Fastnacht, R. Ramesh, and L. H. Chen, Science (1979) **264**, 413 (1994).
[31] Y. Tomioka, X. Z. Yu, T. Ito, Y. Matsui, and Y. Tokura, Phys Rev B **80**, 094406 (2009).
[32] B. A. Gray, S. Middey, G. Conti, A. X. Gray, C.-T. Kuo, A. M. Kaiser, S. Ueda, K. Kobayashi, D. Meyers, M. Kareev, I. C. Tung, J. Liu, C. S. Fadley, J. Chakhalian, and J. W. Freeland, Sci Rep **6**, 33184 (2016).
[33] H. Taimatsu, K. Wada, H. Kaneko, and H. Yamamura, Journal of the American Ceramic Society **75**, 401 (1992).
[34] P. Kaur and K. Singh, Ceram Int **46**, 5521 (2020).
[35] S. P. Jiang, J Mater Sci **43**, 6799 (2008).
[36] T. Lottermoser, M. Fiebig, D. Fröhlich, S. Leute, and K. Kohn, J Magn Magn Mater **226–230**, 1131 (2001).
[37] N. Abramov, V. Chichkov, S. E. Lofland, and Y. M. Mukovskii, J Appl Phys **109**, (2011).
[38] B. Sattibabu, A. K. Bhatnagar, S. Rayaprol, D. Mohan, D. Das, M. Sundararaman, and V. Siruguri, Physica B Condens Matter **448**, 210 (2014).

[39] P. Cheng, W. Sun, Y. Fang, B. Kang, W. Lv, Q. Xiao, J. Zhang, and F. Chen, Ceram Int **46**, 15958 (2020).
[40] B. Lorenz, ISRN Condensed Matter Physics **2013**, 1 (2013).
[41] V. Markovich, E. Rozenberg, G. Gorodetsky, B. Revzin, J. Pelleg, and I. Felner, Phys Rev B **62**, 14186 (2000).
[42] B. Munisha, B. Mishra, and J. Nanda, Journal of Rare Earths **41**, 19 (2023).
[43] C. Leon, J. Santamaria, and B. A. Boukamp, MRS Bull **38**, 1056 (2013).
[44] J. Nordlander, M. D. Rossell, M. Campanini, M. Fiebig, and M. Trassin, Phys Rev Mater **4**, 124403 (2020).
[45] N. V. Tokiy, B. I. Perekrestov, D. L. Savina, and I. A. Danilenko, Physics of the Solid State **53**, 1827 (2011).
[46] C. Ahamer, A. K. Opitz, G. M. Rupp, and J. Fleig, J Electrochem Soc **164**, F790 (2017).
[47] M. Kurumada, H. Hara, and E. Iguchi, Acta Mater **53**, 4839 (2005).
[48] J. E. Bauerle and J. Hrizo, Journal of Physics and Chemistry of Solids **30**, 565 (1969).
[49] R. Pornprasertsuk, P. Ramanarayanan, C. B. Musgrave, and F. B. Prinz, J Appl Phys **98**, (2005).
[50] D. A. Wright, J. S. Thorp, A. Aypar, and H. P. Buckley, J Mater Sci **8**, 876 (1973).
[51] J. S. Thorp and H. P. Buckley, J Mater Sci **8**, 1401 (1973).
[52] J. S. Thorp, A. Aypar, and J. S. Ross, J Mater Sci **7**, 729 (1972).
[53] F. K. Moghadam, T. Yamashita, and D. A. Stevenson, J Mater Sci **18**, 2255 (1983).
[54] X. Xu and W. Wang, Modern Physics Letters B **28**, 1430008 (2014).
[55] K. H. Wu, H.-J. Chen, Y. T. Chen, C. C. Hsieh, C. W. Luo, T. M. Uen, J. Y. Juang, J.-Y. Lin, T. Kobayashi, and M. Gospodinov, EPL (Europhysics Letters) **94**, 27006 (2011).
[56] U. Aschauer, R. Pfenninger, S. M. Selbach, T. Grande, and N. A. Spaldin, Phys Rev B **88**, 054111 (2013).
[57] S.-Y. Choi, S.-D. Kim, M. Choi, H.-S. Lee, J. Ryu, N. Shibata, T. Mizoguchi, E. Tochigi, T. Yamamoto, S.-J. L. Kang, and Y. Ikuhara, Nano Lett **15**, 4129 (2015).
[58] D. S. Aidhy and W. J. Weber, J Mater Res **31**, 2 (2016).
[59] J. R. Petrie, C. Mitra, H. Jeen, W. S. Choi, T. L. Meyer, F. A. Reboredo, J. W. Freeland, G. Eres, and H. N. Lee, Adv Funct Mater **26**, 1564 (2016).
[60] D. S. Aidhy and K. Rawat, J Appl Phys **129**, (2021).
[61] J. Maier, Berichte Der Bunsengesellschaft Für Physikalische Chemie **89**, 355 (1985).
[62] J. Maier, Progress in Solid State Chemistry **23**, 171 (1995).
[63] J.-S. Lee, S. Adams, and J. Maier, J Electrochem Soc **147**, 2407 (2000).
[64] J. Maier, Solid State Ion **131**, 13 (2000).
[65] J. MAIER, Solid State Ion **23**, 59 (1987).
[66] M. Coduri, M. Karlsson, and L. Malavasi, J Mater Chem A Mater **10**, 5082 (2022).
[67] A. D. Poletayev, J. A. Dawson, M. S. Islam, and A. M. Lindenberg, Nat Mater **21**, 1066 (2022).
[68] F. Sánchez-De Jesús, C. A. Cortés-Escobedo, A. M. Bolarín-Miró, A. Lira-Hernández, and G. Torres-Villaseñor, Ceram Int **38**, 2139 (2012).

Supplemental Materials for "**Billion-Fold Enhancement of Room-Temperature Ionic Conductivity in h-RMnO$_3$/YSZ Heterostructures via Electric-Field-Assisted Oxygen Deficiency Engineering**"

Detian Yang,[1] Yaohua Liu,[2] Liang Dai,[1] Zhihang Xu,[1] Xiaoshan Xu[3,4]*

[1] Shanghai Key Laboratory of High Temperature Superconductors, Department of Physics, Shanghai University, Shanghai 200444, China

[2] Second Target Station, Oak Ridge National Laboratory, Oak Ridge, Tennessee 37830, USA

[3] Department of Physics and Astronomy, University of Nebraska, Lincoln, Nebraska 68588, USA44

[4] Nebraska Center for Materials and Nanoscience, University of Nebraska, Lincoln, Nebraska 68588, USA

# 1. X-ray Diffraction (XRD) Structural Characterization of h-$R$MnO$_3$ films

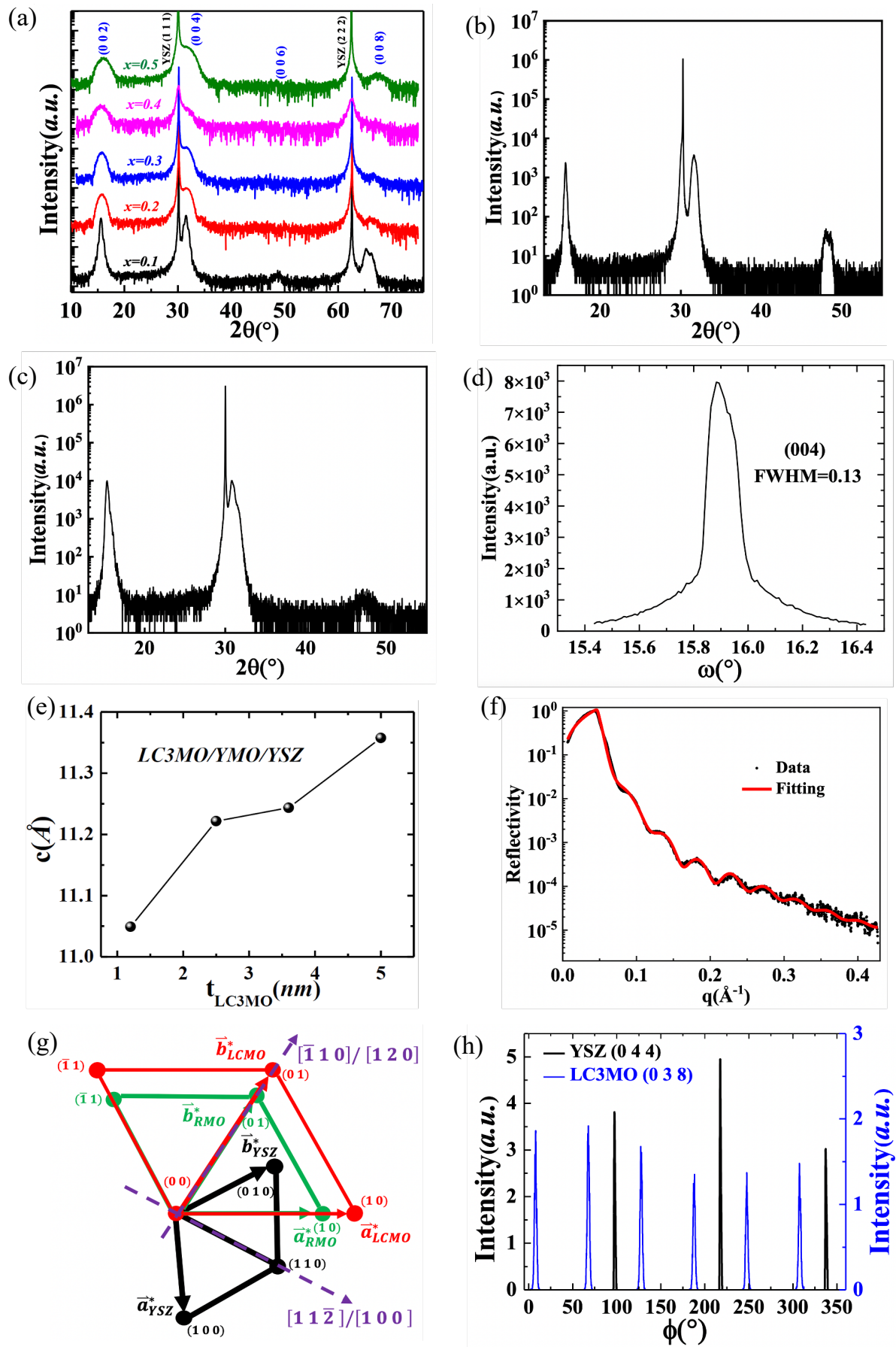

Fig.S1. $\theta - 2\theta$ scans of XRD for $R$MnO$_3$(001)/YSZ (111) films. (a) h-Lu$_{1-x}$Ca$_x$MnO$_3$ (001)/h-YMnO$_3$(001)/YSZ (111); total thicknesses for cases x=0.1, 0.2, 0.3, 0.4, 0.5 are 24 nm, 5 nm, 5nm, 5nm and 4 nm, respectively, while the thickness of the buffer layer h-YMnO$_3$ keeps at 1nm. (b) a 90-nm h-YMnO$_3$(001)/YSZ (111). (c) a 55 nm h-LuMnO$_3$(001)/YSZ (111). (d) rocking curve for (004) peak of a 13 nm h-Lu$_{0.7}$Ca$_{0.3}$MnO$_3$ (001)/h-YMnO$_3$(001)/YSZ (111). (e) Thickness $t_{LC3MO}$ dependent out-of-plane lattice constant c of h-L$_{0.7}$Ca$_{0.3}$MnO$_3$ films grown with the h-YMnO$_3$ buffer layer; the lattice constant value was derived from (0 0 2) peaks of the samples. Here, YMO and LC3MO denote h-YMnO$_3$, and h-Lu$_{0.7}$Ca$_{0.3}$MnO$_3$, respectively. (f) X-ray reflection data (black dots) and its fitting (red line) of a 13 nm h-Lu$_{1-x}$Ca$_x$MnO$_3$ (001)/h-YMnO$_3$(001)/YSZ (111). (g) In-plane reciprocal primitive unit cells for the YSZ substrate (black), h-$R$MnO$_3$ ($R$= Y or Lu$_{1-x}$Ca$_x$, x=0, 0.1, 0.2, 0.3, 0.4, 0.5) (green) and h-Lu$_{1-x}$Ca$_x$MnO$_3$ (red); $a_X^*, b_X^*$ denote the in-plane reciprocal lattice constants for material $X$; purple dashed arrows represent two perpendicular e-beam directions. LCMO stands for Lu$_{1-x}$Ca$_x$MnO$_3$, and $R$MO for $R$MnO$_3$. (h) $\phi$-scan of h-Lu$_{0.7}$Ca$_{0.3}$MnO$_3$ films and the YSZ substrate for a 13 nm h-Lu$_{0.7}$Ca$_{0.3}$MnO$_3$/h-YMnO$_3$/YSZ thin film sample; here LC3MO stands for h-Lu$_{0.7}$Ca$_{0.3}$MnO$_3$.

## 2. Thermal Stability of a 5.6 nm h-Lu$_{0.7}$Ca$_{0.3}$MnO$_3$ (001)/h-YMnO$_3$(001)/YSZ (111)

To check the thermal stability of Ca-doped hexagonal manganite films, we annealed a 5.6 nm h-Lu$_{0.7}$Ca$_{0.3}$MnO$_3$ (001)/h-YMnO$_3$(001)/YSZ (111) sample in a sequence of temperatures from 700-1050 °C in the atmosphere of O$_2$ and measured its $\theta - 2\theta$ scans of XRD after each run. As indicated in Fig.S2, new unknow phases have not appeared until 1000 °C, which appears presumably as a result of the decomposition of Lu$_{0.7}$Ca$_{0.3}$MnO$_3$.

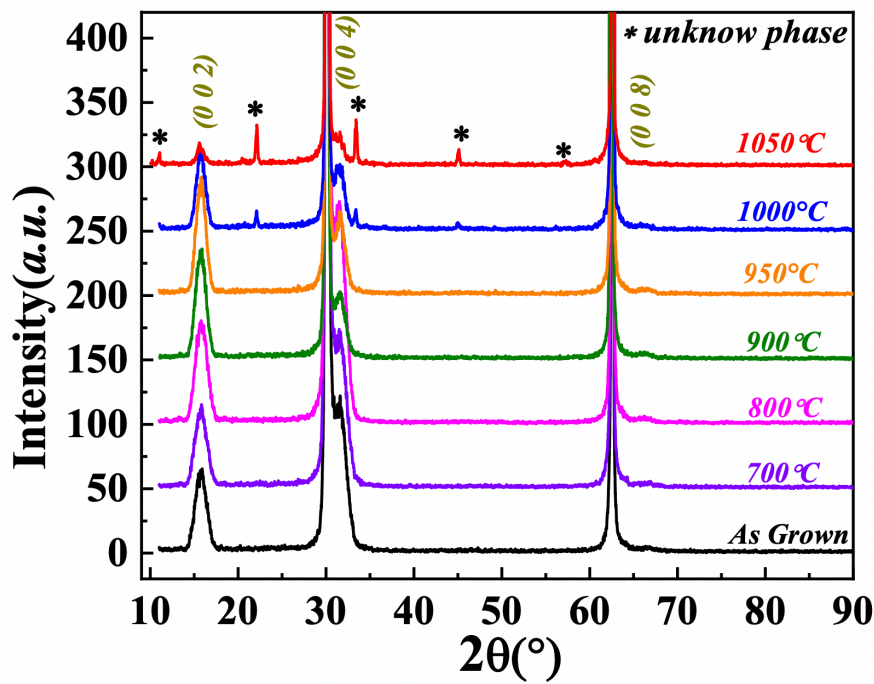

Fig.S2. Annealing effect on the crystal structure of h-Lu$_{0.7}$Ca$_{0.3}$MnO$_3$(001)/h-YMnO$_3$(001)/YSZ (111). A series of $\theta - 2\theta$ scans of XRD was carried out after the sample was removed from the PLD growth chamber and/or then annealed in a furnace with varying temperatures in the oxygen atmosphere.

## 3. Low-temperature R-T curve of a 13 nm h-Lu$_{0.7}$Ca$_{0.3}$MnO$_3$(001)/h-YMnO$_3$(001) /YSZ (111) thin film sample

The low-temperature (20<T<300K) R-T characterization of a h-Lu$_{0.7}$Ca$_{0.3}$MnO$_3$(001)/h-YMnO$_3$(001)/YSZ (111) thin film sample was conducted, and the result was plotted in Fig.S3. It demonstrates a feature of extrinsic semiconductors with a plateau in the range 20<T<150K.

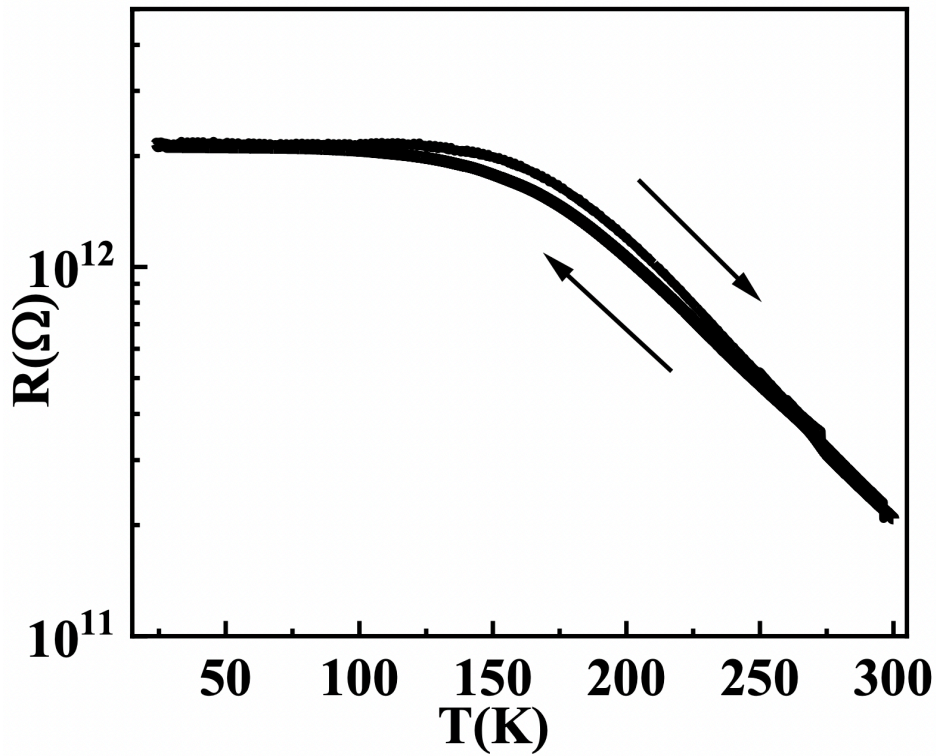

**Fig.S3**. R-T curve of a 13 nm h-Lu$_{0.7}$Ca$_{0.3}$MnO$_3$(001)/h-YMnO$_3$(001)/YSZ(111) thin film sample measured in the temperature range 20K<T<300K and in the vacuum of 0.1Pa. Black arrows indicate the measurement directions which were done first in the cooling process and then in the heating process.

## 4. Arrhenius-type relation fittings of R-T curves of h-Lu$_{0.7}$Ca$_{0.3}$MnO$_3$(001)/h-YMnO$_3$(001)/YSZ (111) heterostructures with varying film thicknesses

In a single crystal ionic conductor like YSZ, the temperature dependent conductivity is given by Arrhenius-type relation [1]

$$\sigma = \frac{\sigma_0}{k_B T} e^{(-\frac{E_a}{k_B T})} \qquad (1)$$

where $k_B$ is the Boltzmann constant, $\sigma_0$ a constant and the activation energy of vacancies $E_a = E_m + E_{va}$ include two parts: $E_m$ is the energy of vacancy migration and $E_{va}$ the energy of vacancy association. Here we rewrite (1) in terms of resistance

$$R = R_0 T e^{(\frac{E_a}{k_B T})} \qquad (2)$$

For the R-T data of our samples, as shown in Fig.S4(a-e.1)-(a-e.2), the range between the "dip" and the downward or jump points measured in the heating process((a.1)-(e.1)) and the cooling branches of R-T curves((a.2)-(e.2)) could be well fitted by the formula (2), although in the latter case, the range is divided into two segments and each of them was fitted with different parameters $R_0$ and $E_A$, reflecting the changes in the properties of charge carriers as the temperature T decreases.

In Fig.S4(a.1)-(e.1), the fitted activation energies are close to the activation energy of YSZ [2-4] as shown in Fig.2(a) and Fig.S4(f.1), indicating that the oxygen vacancies of YSZ dominate the charge carriers the range between the "dip" and the downward or jump points. Besides, the activation energies for the cooling process demonstrate a dramatic decrease once the film thickness surpasses the critical thickness, as illustrated in Fig.S4(f.2). This shows that the high conductivity state after the sharp resistance drop is a result of great enhancement of the mobility of charge carriers.

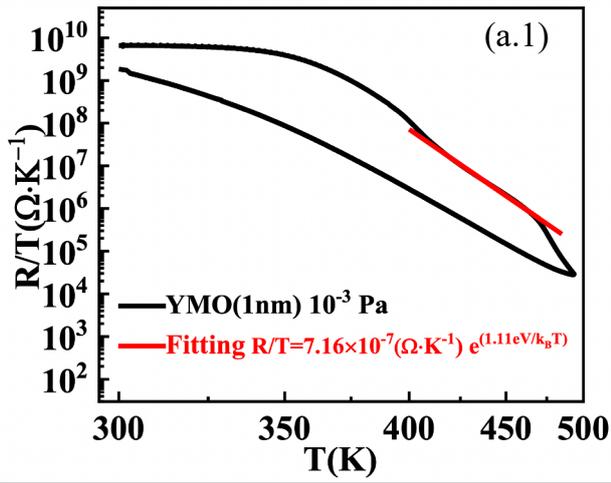
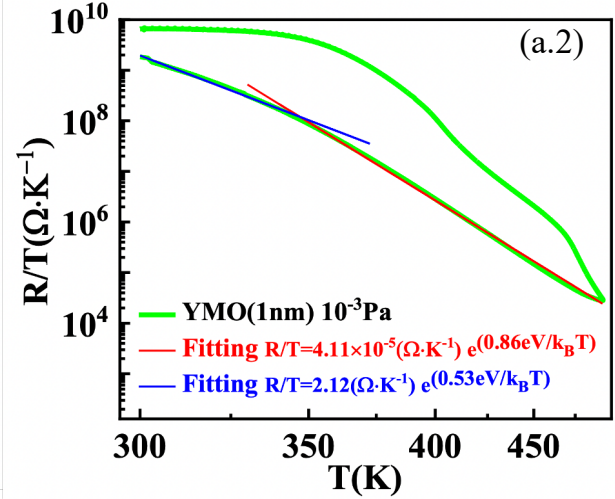
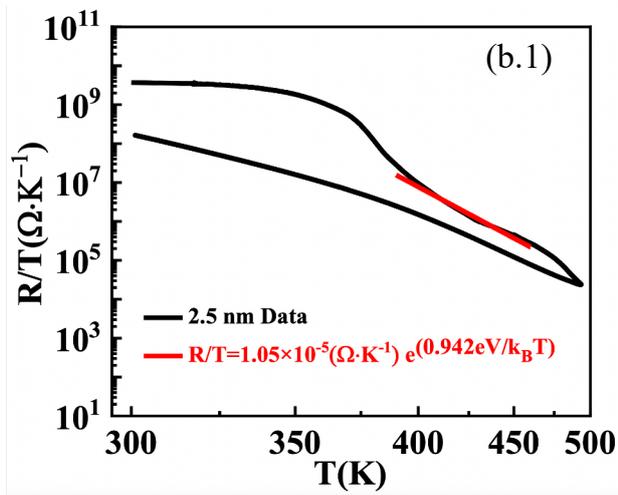
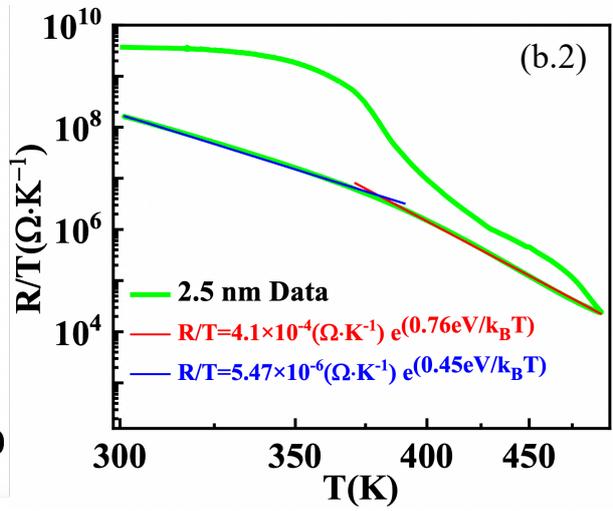
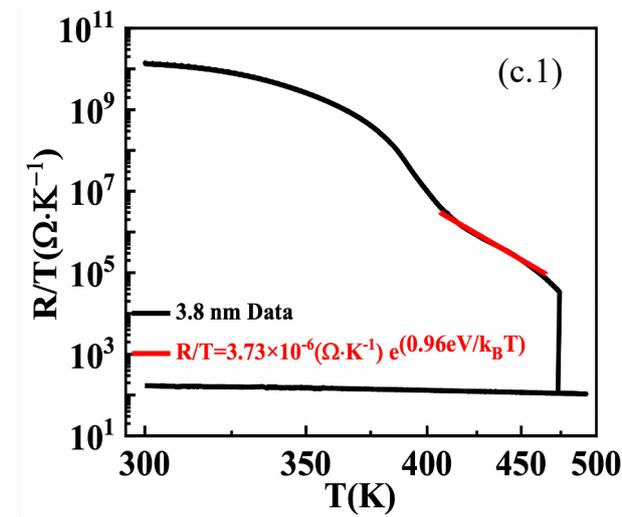
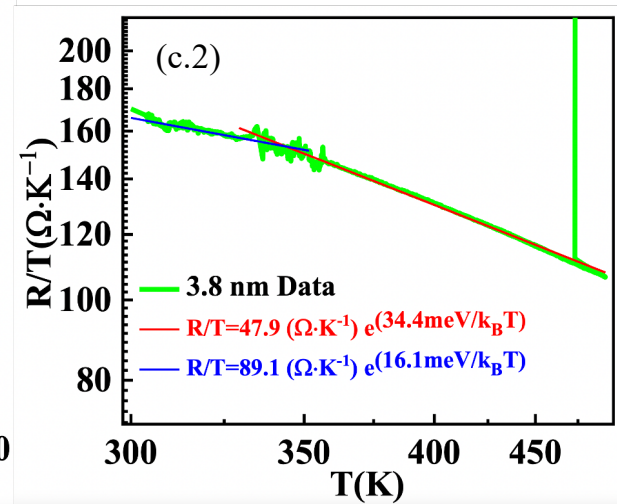

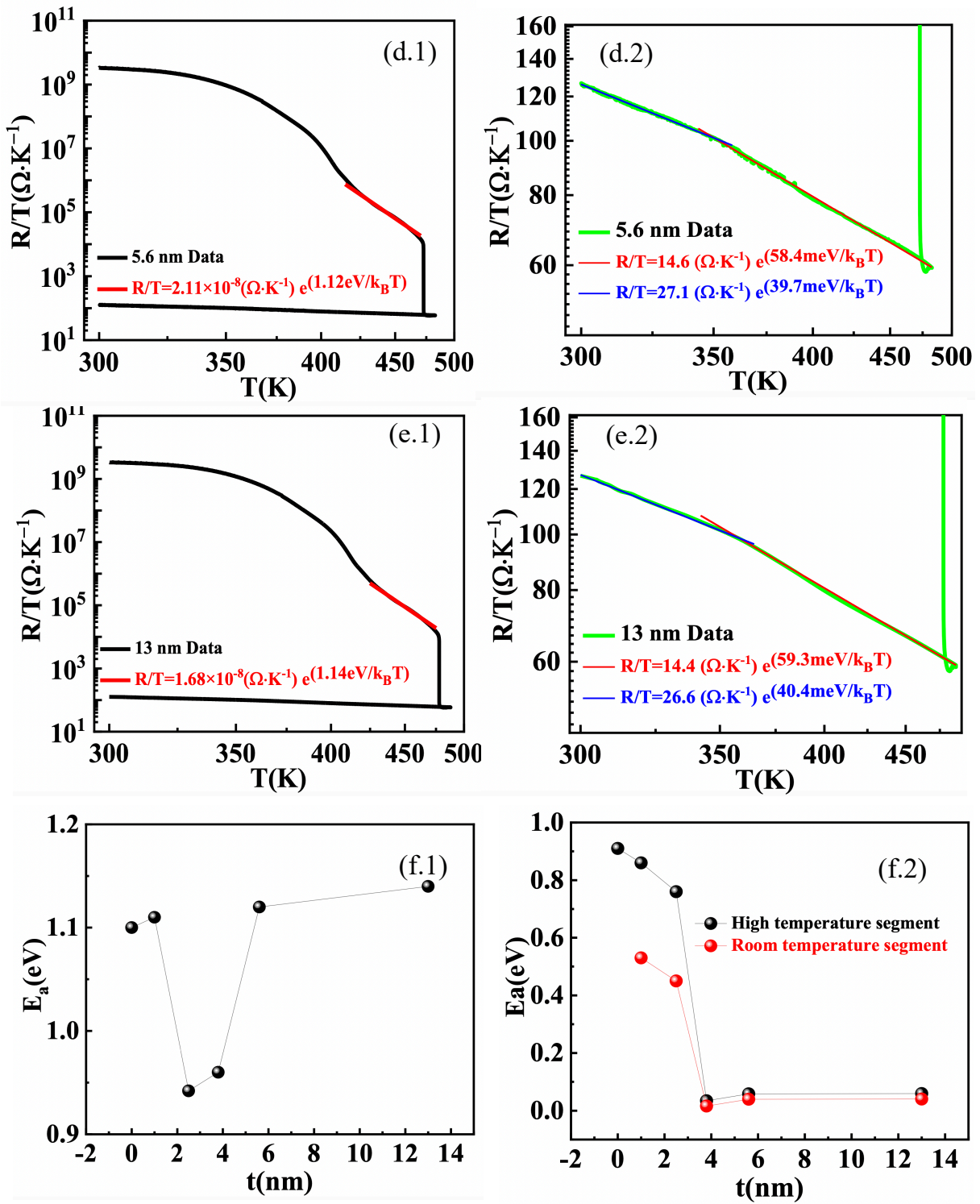

**Fig.S4**. Arrhenius-type relation fittings for R-T curves. **(a)**1 nm h-YMnO$_3$(001) (blue curve: data; red line: fitting); **(b)**2.5 nm h-Lu$_{0.7}$Ca$_{0.3}$MnO$_3$(001)/h-YMnO$_3$(001);**(c)** 3.8 nm h-Lu$_{0.7}$Ca$_{0.3}$MnO$_3$(001)/h-YMnO$_3$(001);**(d)** 5.6 nm h-Lu$_{0.7}$Ca$_{0.3}$MnO$_3$(001)/h-YMnO$_3$(001);**(e)** 13 nm h-Lu$_{0.7}$Ca$_{0.3}$MnO$_3$(001)/h-YMnO$_3$(001). **(a.1)-(e.1)** The range between the "dip" and downturn or jump point

in the heating-up R-T curves (see the main text); the data and the fittings were plotted with black and red lines, respectively. **(a.2)-(e.2)** The cooling-down R-T curves; green lines denote the data, while red and blue lines indicate two fittings. **(f.1)** Thickness dependent activation energy $E_a$-$t$ extracted from fittings in (a.1)-(e.1). **(f.2)** Thickness dependent activation energy $E_a$-$t$ extracted from fittings in (a.2)-(e.2). Black (red) balls denote the activation energies of the high-temperature (room-temperature) segments with fitting lines in red (blue) in (a.2)-(e.2). In (f), the data for YSZ have been added for comparison.

## 5. Structural stability in R-T measurements

To ascertain the potential structural changes in the R-T measurement process, we measured the XRD data of a 5.6 nm h-Lu$_{0.7}$Ca$_{0.3}$MnO$_3$/h-YMnO$_3$/YSZ (Fig.S5(a)) and a 55 nm h-LuMnO$_3$/YSZ (Fig.S5(b)) before and after the R-T characterization. As shown in Fig.S5(a.1) and (b.1), no substantial changes were observed in the $\theta - 2\theta$ scans, indicating the overall crystal structures of manganite film(hexagonal) and YSZ substrate (cubic) are preserved. In Fig.S5(a.1), also shown by the blue curve is the $\theta - 2\theta$ scan for YSZ (111) after the black sample (induced by R-T measurement as elaborated in the main text) was annealed at 500 for 3 hours. However, the rocking curves were broadened, indeed, as illustrated by the case of (004) peak in Fig.S5(a.2) and (b.2).

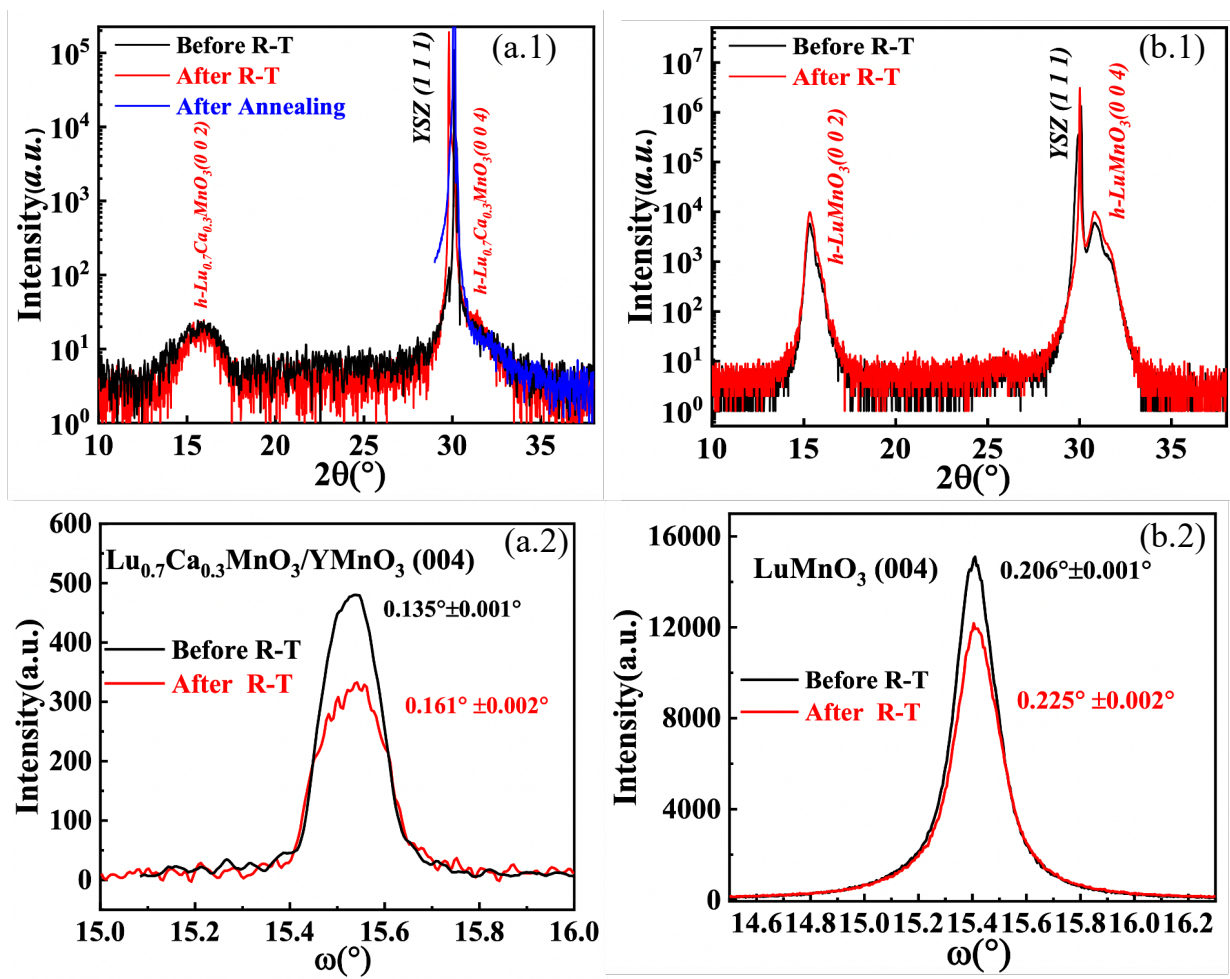

**Fig.S5**. XRD for **(a)** a 5.6 nm h-Lu$_{0.7}$Ca$_{0.3}$MnO$_3$(001)/h-YMnO$_3$(001)/YSZ(111) and **(b)** a 55 nm h-LuMnO$_3$(001)/YSZ(111). **(a.1)-(b.1)** $\theta - 2\theta$ scans measured before (black curve) and after (red curve ) the R-T measurement and after further 3-hour annealing at 500°C in air(blue curve). **(a.2)-(b.2)** Rocking curves for (004) peak measured before (black curve) and after (red curve) the R-T measurement. In (a.2) and (b.2), the full widths at half maximum for rocking curves are labeled by numbers of matching colors with their corresponding curves.

## 6. Absorption spectra of YSZ substrates

As elaborated in the main text, the YSZ substrate gradually transformed to a deep black color from the "dip" to the resistance jump point during the R-T characterization process. To verify this observation, we removed the manganite film using reactive ion etching after completing the R-T measurements. Subsequently, we measured the transmittance by a Stellar Net spectrometer and calculated the absorptance of the black YSZ substrate in the visible light spectrum, plotting these results alongside the absorptance spectrum of a pristine YSZ substrate. As shown in Fig. S6, the black YSZ exhibits an absorptance close to 100% across the visible light range compared to the pristine YSZ, similar to the result seen in the current-blackened 8 mol% YSZ[5] . Note that the unusually high absorptance [6] (>0.5) of the pristine YSZ can be attributed to the reflection from its unpolished side, given our transmittance measurement setup and one-side polished substrate.

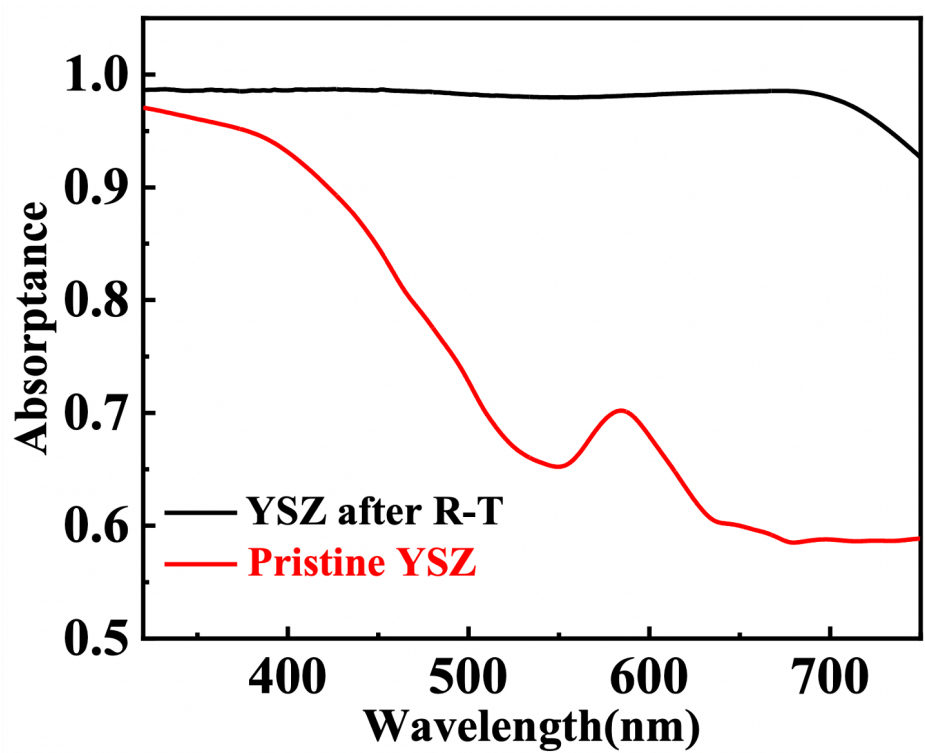

**Fig.S6**. Absorptance spectra for a pristine YSZ substrate and one that went through the temperature-dependent resistance measurement.

## 7. The degradation of conductivity of the 5.6 nm h-Lu$_{0.7}$Ca$_{0.3}$MnO$_3$(001)/h-YMnO$_3$(001)/YSZ (111) thin film sample in reaction to oxygen atmosphere.

After the R-T measurement shown in Fig.2(b), at 300K, we gradually exposed the 5.6 nm h-Lu$_{0.7}$Ca$_{0.3}$MnO$_3$(001)/h-YMnO$_3$(001)/YSZ (111) thin film sample to oxygen and continuously measured its current-voltage (I-U) curve in the voltage range -100-100V. As shown in Fig.S7, the I-U curve shifts from the low resistance state to higher resistance states in reaction to oxygen atmosphere, wherein, a sharp transition can be seen as the oxygen pressure turns from 5 Pa to $5\times10^4$ Pa.

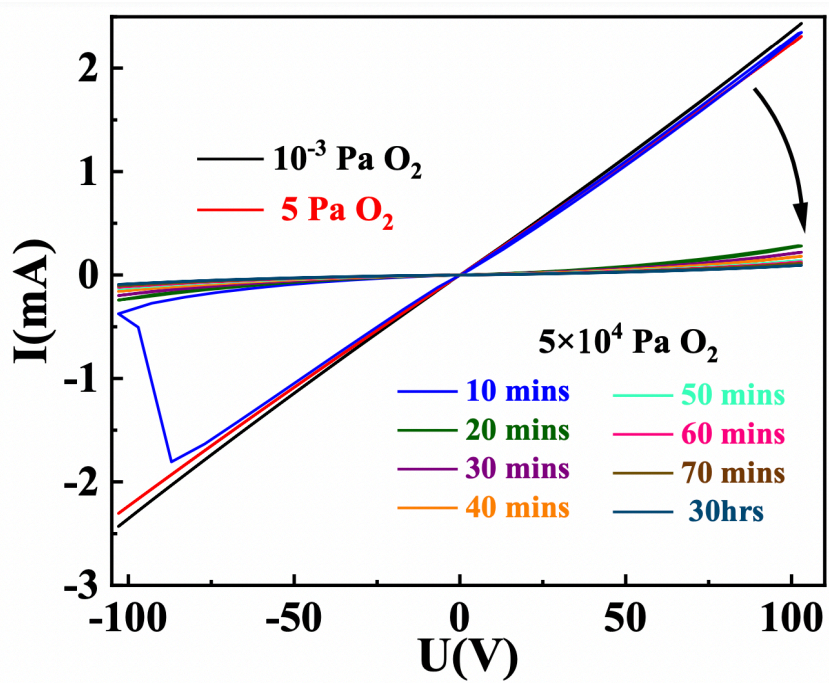

**Fig.S7**. I-U curves of the 5.6 nm h-Lu$_{0.7}$Ca$_{0.3}$MnO$_3$(001)/h-YMnO$_3$(001)/YSZ (111) thin film sample measured at 300K with varying oxygen pressures after the R-T measurement shown in Fig.2(b).

## 8. More R-T Curves of h-*R*MnO$_3$/YSZ heterostructures

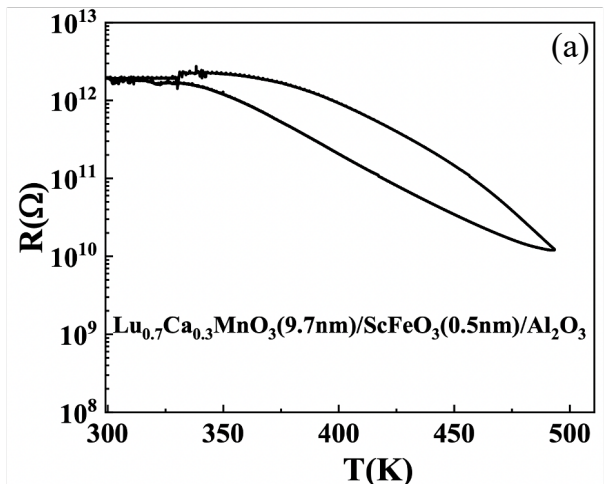
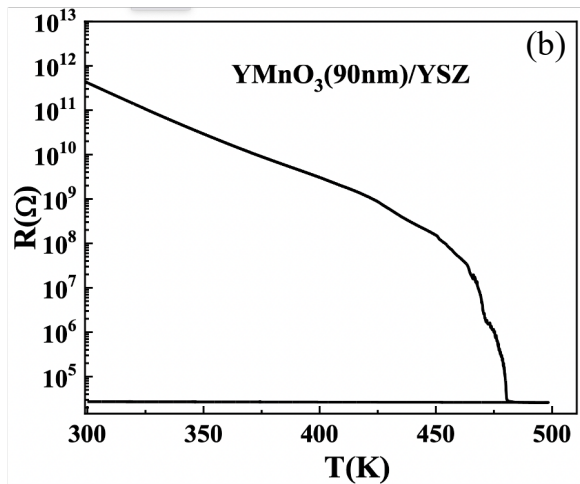
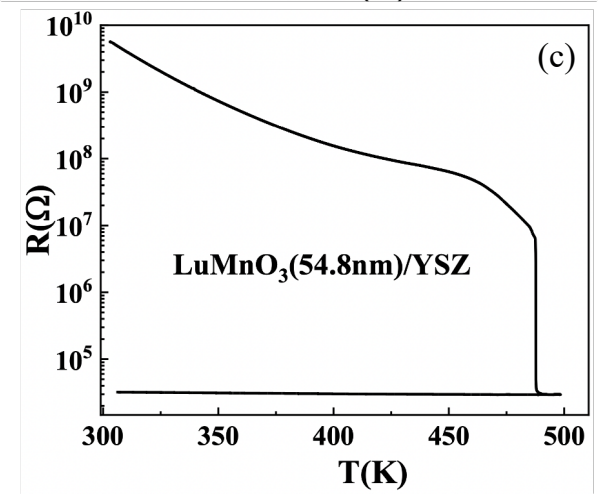
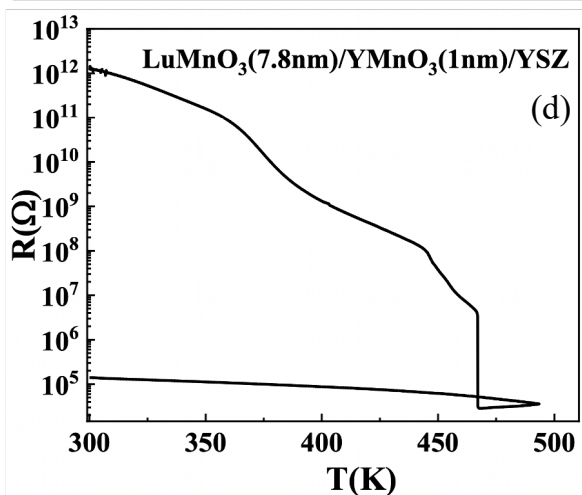
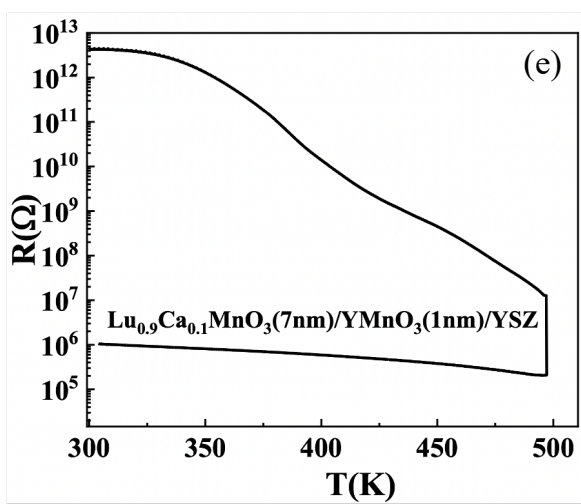
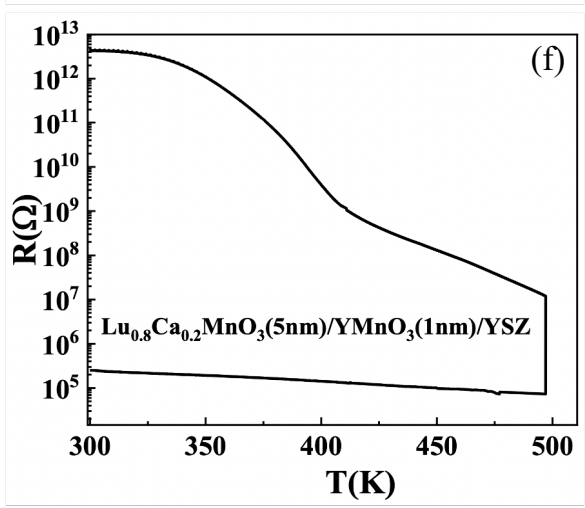

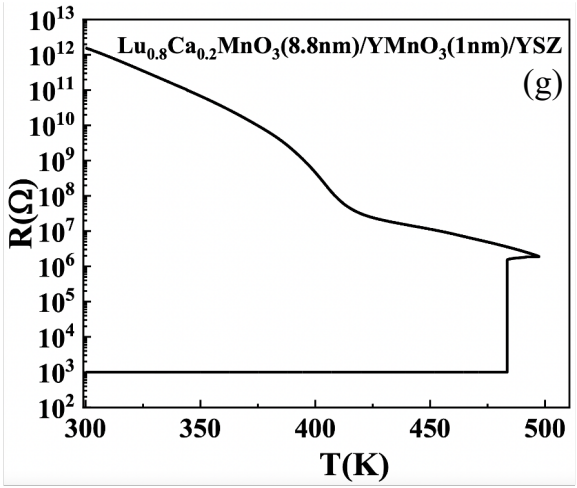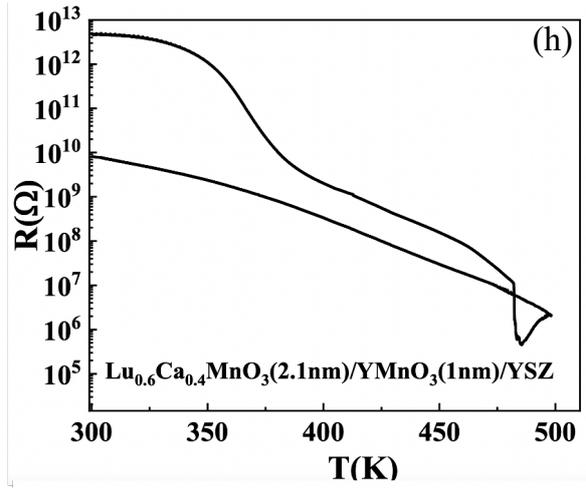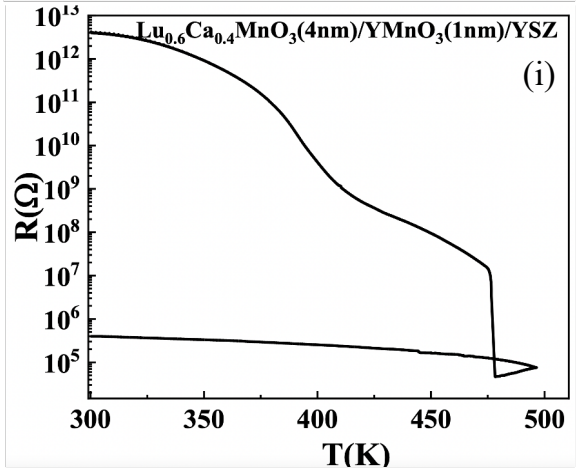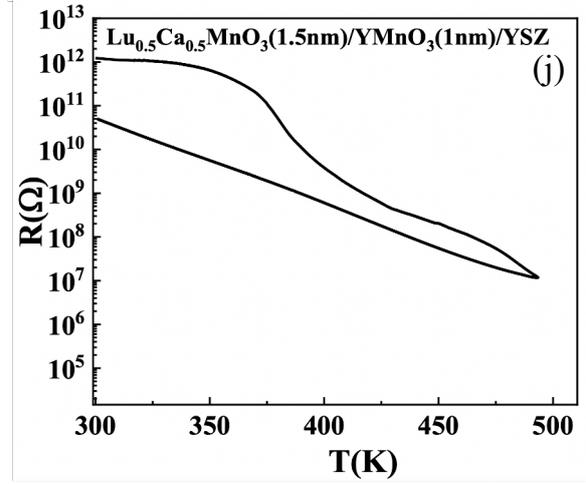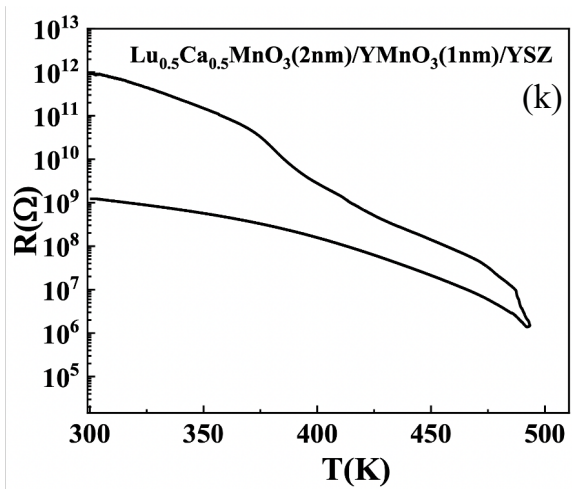

**Fig.S8**. More R-T curves for h-$R$MnO$_3$ thin film heterostructures:

(a) h-Lu$_{0.7}$Ca$_{0.3}$MnO$_3$(9.7nm)/h-ScFeO$_3$(0.5nm)/Al$_2$O$_3$(0001);

(b) h-YMnO$_3$(90nm)/YSZ(111);

(c) h-LuMnO$_3$(54.8nm)/YSZ(111);

(d) h-LuMnO$_3$(7.8nm)/h-YMnO$_3$(1nm)/YSZ(111);

(e) h-Lu$_{0.9}$Ca$_{0.1}$MnO$_3$(7nm)/h-YMnO$_3$(1nm)/YSZ(111);

(f) h-Lu$_{0.8}$Ca$_{0.2}$MnO$_3$(5nm)/h-YMnO$_3$(1nm)/YSZ(111);

(g) h-Lu$_{0.8}$Ca$_{0.2}$MnO$_3$(8.8nm)/h-YMnO$_3$(1nm)/YSZ(111);

(h) h-Lu$_{0.6}$Ca$_{0.4}$MnO$_3$(2.1nm)/h-YMnO$_3$(1nm)/YSZ(111);

(i) h-Lu$_{0.6}$Ca$_{0.4}$MnO$_3$(4nm)/h-YMnO$_3$(1nm)/YSZ(111);

(j) h-Lu$_{0.5}$Ca$_{0.5}$MnO$_3$(1.5nm)/h-YMnO$_3$(1nm)/YSZ(111);

(k) h-Lu$_{0.5}$Ca$_{0.5}$MnO$_3$(2nm)/h-YMnO$_3$(1nm)/YSZ(111).

**References:**

[1] J. Luo, D.P. Almond, R. Stevens, J. Am. Ceram. Soc. 83, 1703 (2000)

[2] M. Kurumada, H. Hara, and E. Iguchi, Acta Mater 53, 4839 (2005).

[3] J. E. Bauerle and J. Hrizo, Journal of Physics and Chemistry of Solids 30, 565 (1969).

[4] R. Pornprasertsuk, P. Ramanarayanan, C. B. Musgrave, and F. B. Prinz, J Appl Phys 98, (2005).

[5] D. A. Wright, J. S. Thorp, A. Aypar, and H. P. Buckley, J Mater Sci **8**, 876 (1973).

[6] Kunz, M., Kretschmann, H., Assmus, W., & Klingshirn, C. *Journal of Luminescence*, *37*(3), 123–131 (1987).